\documentclass[
    ,final            
  ]
  {aipproc}

\usepackage{amssymb}

\layoutstyle{8x11single}

\def\be{\begin{equation}}
\def\ee{\end{equation}}
\def\bea{\begin{eqnarray}}
\def\eea{\end{eqnarray}}
\def\bear{\begin{array}}
\def\ear{\end{array}}
\def\bfig{\begin{figure}}
\def\efig{\end{figure}}
\def\bcen{\begin{center}}
\def\ecen{\end{center}}
\def\bi{\begin{itemize}}
\def\ei{\end{itemize}}

\def\raw{\rightarrow}

\def\slashchar#1{\setbox0=\hbox{$#1$}
   \dimen0=\wd0 \setbox1=\hbox{/} \dimen1=\wd1
   \ifdim\dimen0>\dimen1 \rlap{\hbox to \dimen0{\hfil/\hfil}} #1
   \else  \rlap{\hbox to \dimen1{\hfil$#1$\hfil}} / \fi}

\begin{document}

\title{Single photon production induced by (anti)neutrino neutral current scattering on nucleons and nuclear targets}


\author{L. Alvarez-Ruso}{
  address={Instituto de F\'isica Corpuscular (IFIC), Centro Mixto CSIC-Universidad de Valencia, \\Institutos de Investigaci\'on de Paterna, E-46071 Valencia, Spain}
}

\author{J. Nieves}{
  address={Instituto de F\'isica Corpuscular (IFIC), Centro Mixto CSIC-Universidad de Valencia, \\Institutos de Investigaci\'on de Paterna, E-46071 Valencia, Spain}
}

\author{E. Wang}{
  address={Instituto de F\'isica Corpuscular (IFIC), Centro Mixto CSIC-Universidad de Valencia, \\Institutos de Investigaci\'on de Paterna, E-46071 Valencia, Spain}, 
altaddress={Department of Physics, Zhengzhou University, Zhengzhou, Henan 450001, China}
}


\begin{abstract}
We review our theoretical approach to neutral current photon emission on nucleons and nuclei in the few-GeV energy region, relevant for neutrino oscillation experiments. These reactions are dominated by the weak excitation of the $\Delta(1232)$ resonance but there are also important non-resonant contributions.  We have also included terms mediated by nucleon
excitations from the second resonance region. On nuclei, Pauli blocking, Fermi motion and the in-medium $\Delta$ resonance broadening have been taken into account for both incoherent and coherent reaction channels. With this model, the number and distributions of photon events at the MiniBooNE and T2K experiments have been obtained. We have also compared to the NOMAD upper limit at higher energies. The implications of our findings and future perspectives are discussed. 
\end{abstract}

\maketitle


\section{Introduction}

A good understanding of (anti)neutrino cross sections is crucial to reduce systematic uncertainties in oscillation experiments~\cite{Garvey:2014exa}. Our present knowledge  of neutrino-nucleus interactions has been improved by a new generation of oscillation and cross section experiments. Over the last decade, K2K, NOMAD, MiniBooNE, SciBooNE, MINOS, and more recently T2K and MINER$\nu$A have obtained a wealth of data on quasielastic(-like) scattering, incoherent and coherent single pion production, and inclusive cross sections.  These results challenge our understanding of neutrino interactions with matter and have triggered a renewed theoretical interest~\cite{Alvarez-Ruso:2014bla}. 

One of the possible reaction channels is photon emission induced by neutral current (NC) interactions (NC$\gamma$), which can occur on single nucleons and on nuclear targets  with incoherent or  coherent reaction mechanisms.  Weak photon emission has a small cross section compared, for example, with pion production, the most important inelastic reaction channel.  In spite of this, NC photon emission turns out to be one of the largest backgrounds in $\nu_\mu \to \nu_e$ $(\bar\nu_\mu \to \bar \nu_e)$ oscillation experiments where electromagnetic showers from electrons (positrons) and photons are not distinguishable. 

The first effort to put the description of NC photon emission on solid theoretical grounds was reported in Ref.~\cite{Hill:2009ek}. The reaction on nucleons was studied with a microscopic model developed in terms of hadronic degrees of freedom: nucleon, $\Delta(1232)$ resonance and mesons. Coherent photon emission off nuclear targets was also evaluated, treating the nucleus as a scalar particle and introducing a form factor to ensure that the coherence is restricted to low-momentum transfers. The NC$\gamma$ reactions on nucleons and nuclei were also studied using a chiral effective field theory of nuclei~\cite{Serot:2012rd,Zhang:2012aka,Zhang:2012xi}, phenomenologically extended to the intermediate energies ($E_\nu \sim 1$~GeV) in Ref.~\cite{Zhang:2012xn}.  In this model, a rather strong in-medium suppression of the  $\Delta(1232)$ excitation is compensated by rapidly growing contact terms. 

Our theoretical model~\cite{Wang:2013wva}, summarized below, extends and improves several relevant aspects of the existing descriptions. The energy and target mass dependence of the integrated cross sections are displayed and discussed. The model is then applied to the specific conditions of the MiniBooNE, T2K and NOMAD experiments.

\section{Formalism}
\label{sec:formalism}

\subsection{NC$\gamma$ on nucleons}

The cross section for 
\be
\nu,\bar\nu (k) +\,  N (p) \to  \nu,\bar\nu (k') +\, N (p') +\, \gamma (k_\gamma)
\ee
in the laboratory frame is given by
\begin{equation}
\frac{d^{\,3}\sigma_{(\nu,\bar\nu)}}{dE_\gamma d\Omega(\hat{k}_\gamma)} =
    \frac{E_\gamma}{ |\vec{k}|}\frac{G^2}{16\pi^2}
     \int  \frac{d^3k'}{|\vec{k}^{\prime}\,|}
    L_{\mu\sigma}^{(\nu,\bar\nu)}W^{\mu\sigma}_{{\rm NC}\gamma} \label{eq:sec} \,,
\end{equation}
in terms of contraction of the leptonic tensor 
\be
L_{\mu\sigma}^{(\nu,\bar\nu)} =
 k^\prime_\mu k_\sigma +k^\prime_\sigma k_\mu
+ g_{\mu\sigma} \frac{q^2}2 \pm i
\epsilon_{\mu\sigma\alpha\beta}k^{\prime\alpha}k^\beta \,,
\ee
with the hadronic one
\begin{eqnarray}
W^{\mu\sigma}_{{\rm NC} \gamma} &=& \frac{1}{4M}\overline{\sum_{\rm
 spins}} \int\frac{d^3p^\prime}{(2\pi)^3} \frac{1}{2E^\prime_N}
  \delta^4(p^\prime+k_\gamma-q-p) \langle N \gamma |
 j^\mu_{\rm NC\gamma}(0) | N \rangle \langle N \gamma | j^\sigma_{\rm NC \gamma}(0) | N
 \rangle^* \label{eq:wmunu-nucleon} \,.
\end{eqnarray}
In these equations, $M$ denotes the nucleon mass, $E_\gamma$ and $E^\prime_N$ are the photon and final nucleon laboratory energies and $q = k -k'$ is the 4-momentum transfer. 
The model is defined by the set of Feynman diagrams for the hadronic current~\cite{Wang:2013wva}.
\begin{equation}
\langle N \gamma |
 j^\mu_{\rm NC\gamma}(0) | N \rangle = \bar u(p') \Gamma^{\mu \rho} u (p) \epsilon^*_\rho(k_\gamma) \,,
\end{equation}
shown in Fig.~\ref{fig:diags}; $\epsilon(k_\gamma)$ is the photon polarization vector.

\bfig[h!]
\includegraphics[width=0.24\textwidth]{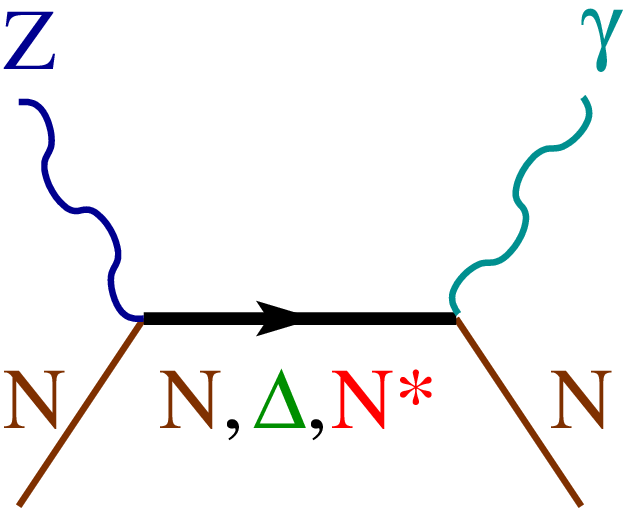}
\hspace{.05\textwidth} 
\includegraphics[width=0.24\textwidth]{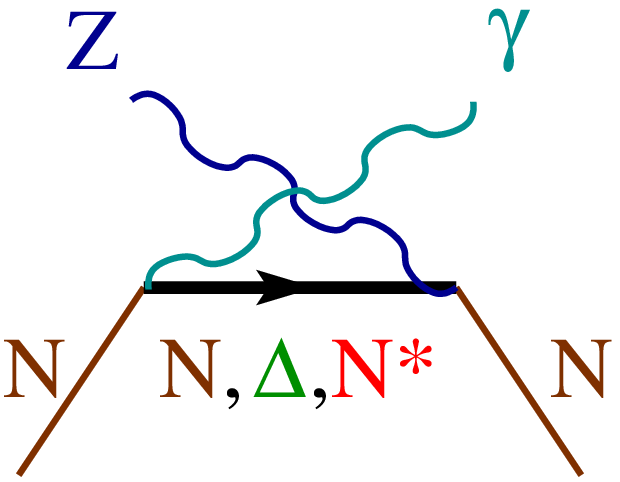}
\hspace{.05\textwidth} 
\includegraphics[width=0.20\textwidth]{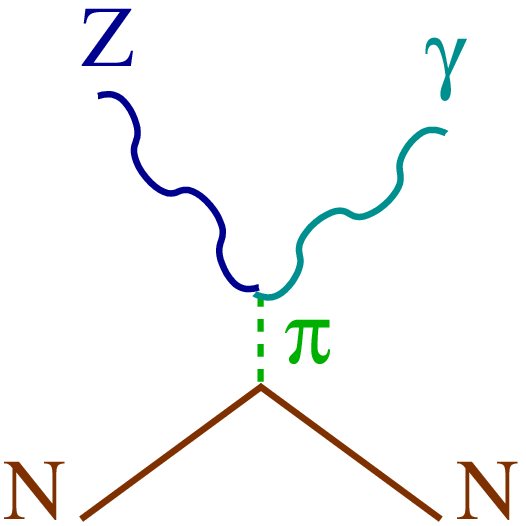}
\caption{\label{fig:diags}  Feynman diagrams for the hadronic current of NC photon emission considered in Ref.~\cite{Wang:2013wva}. The first two diagrams stand for direct and crossed baryon pole terms with nucleons and resonances in the intermediate state: $BP$ and $CBP$ with $B=N$, $\Delta(1232)$, $N(1440)$, $N(1520)$, $N(1535)$. The third diagram represents the $t$-channel pion exchange: $\pi Ex$.}
\efig

The amputated amplitude $\Gamma^{\mu \rho}_{N}$ for nucleon pole terms, $NP$ and $CNP$, is given by
\be
\Gamma^{\mu \rho}_{N} = \Gamma^{\mu \rho}_{NP} + \Gamma^{\mu
  \rho}_{CNP} = i e\,
J^\rho_{EM}(-k_\gamma)\frac{\slashchar{p}+\slashchar{q}+M}{(p+q)^2-M^2
+i\epsilon} J^\mu_{NC}(q)   
 +
i e\,  J^\mu_{NC}(q)\frac{(\slashchar{p}'-\slashchar{q}+M)}{(p'-q)^2-M^2+i\epsilon} 
J^\rho_{EM}(-k_\gamma) \label{eq:NP-CNP} \,. 
\ee
with currents 
\begin{eqnarray}
J^\mu_{NC}(q)&=&\gamma^\mu \tilde{F}_1(q^2)+\frac{i}{2M}\sigma^{\mu\beta}q_\beta \tilde{F}_2(q^2) - \gamma^\mu \gamma_5 \tilde{F}_A(q^2),   \\
J^\mu_{EM}(k_\gamma)&=& \gamma^\mu F_1(0) + \frac{i}{2M}\sigma^{\mu\nu}
(k_{\gamma})_\nu F_2(0)\,.
\end{eqnarray}
At threshold, this mechanism is fully constrained by gauge and chiral symmetries, and the partial conservation of the axial current (PCAC). These terms are infrared divergent when the photon energy $E_\gamma \raw 0$ but this doesn't have practical consequences because very soft photons will not be experimentally resolved. The extension towards higher energy transfers required to make predictions at $E_\nu \sim 1$~GeV is performed using phenomenological parametrizations of the weak form factors. Isospin symmetry allows to relate the  NC vector form factors 
\begin{eqnarray}
 \tilde{F}^{(p)}_{1,2}&=&(1-4 \sin^2\theta_W) F^{(p)}_{1,2} -F^{(n)}_{1,2} -F^{(s)}_{1,2}\\
 \tilde{F}^{(n)}_{1,2}&=&(1-4 \sin^2\theta_W) F^{(n)}_{1,2} -F^{(p)}_{1,2} -F^{(s)}_{1,2} \,,
\end{eqnarray}
to the electromagnetic ones that have been extracted from electron scattering data. For the axial form factors 
\begin{equation}
 \tilde{F}^{(p,n)}_A =  \pm F_A  - F^{(s)}_A\,,
\end{equation}
we have adopted a conventional dipole parametrization~\cite{Bodek:2007ym}. Strange form factors, whose present values are consistent with zero, have been neglected. 

 The $\Delta P$ and $C\Delta P$ terms 
\bea
\Gamma^{\mu \rho}_{\Delta} = \Gamma^{\mu \rho}_{\Delta P} + \Gamma^{\mu
  \rho}_{C\Delta P} &=& i e\,
\gamma^0 \left[ J^{\alpha \rho}_{EM}(p',k_\gamma)\right]^\dagger
\gamma^0\frac{P_{\alpha\beta}
  (p+q)}{(p+q)^2-M_\Delta^2
+i M_\Delta \Gamma_\Delta} J^{\beta \mu}_{NC}(p,q) \nonumber \\   
&+&
i e\, \gamma^0 \left[  J^{\alpha \mu}_{NC}(p',-q) \right]^\dagger
\gamma^0 \frac{P_{\alpha\beta}
  (p^{\,\prime}-q)}{(p^{\,\prime}-q)^2-M_\Delta^2
+i \epsilon}J^{\beta \rho}_{EM}(p,-k_\gamma)\,, \label{eq:DP-CDP}
\eea
with nucleon-$\Delta(1232)$ transition currents 
\begin{eqnarray}
\frac12 J^{\beta\mu}_{NC}(p,q) &=& \left[ \frac{\tilde{C}^V_3(q^2)}{M} (g^{\beta\mu} \slashchar{q}
    -q^\beta \gamma^\mu ) +\frac{\tilde{C}^V_4(q^2)}{M^2} (g^{\beta\mu} q \cdot p_\Delta
    -q^\beta p^\mu_\Delta )+\frac{\tilde{C}^V_5(q^2)}{M^2} (g^{\beta\mu} q \cdot p
    -q^\beta p^\mu ) \right] \gamma_5 \nonumber\\
 &+&\frac{\tilde{C}^A_3(q^2)}{M} (g^{\beta\mu}\slashchar{q}-q^\beta \gamma^\mu ) 
   + \frac{\tilde{C}^A_4(q^2)}{M^2} (g^{\beta\mu} q \cdot p_\Delta
    -q^\beta p^\mu_\Delta ) + \tilde{C}^A_5(q^2) g^{\beta\mu}\label{eq:nc}\,,
\end{eqnarray}
\begin{eqnarray}
J^{\beta\rho}_{EM}(p,-k_\gamma) &=& -\left[ \frac{C^V_3(0)}{M}
  (g^{\beta\rho} \slashchar{k}_\gamma-k_\gamma^{\beta} \gamma^\rho )
  +\frac{C^V_4(0)}{M^2} (g^{\beta\rho} k_\gamma \cdot p_{\Delta c}
  -k_\gamma^\beta p^\rho_{\Delta c} )
+ \frac{C^V_5(0)}{M^2} (g^{\beta\rho} k_\gamma \cdot p -k_\gamma^\beta p^\rho ) \right] \gamma_5 \,,
\end{eqnarray}
is given in terms of vector 
\begin{equation}
\tilde{C}^V_{3-5}(q^2) = (1-2 \sin^2 \theta_W) C^V_{3-5}(q^2)
\end{equation}
and axial ($\tilde{C}^A_{3-5}=C^A_{3-5}$) form factors. The vector form factors are related to the helicity amplitudes extracted in the analysis of pion photo- and electro-production data. We have adopted the parametrizations of the helicity amplitudes obtained with the unitary isobar model MAID~\cite{Drechsel:2007if}. Assuming the so called Adler model~\cite{Adler:1968tw,Bijtebier:1970ku}, the axial current depends on just one form factor
\begin{equation}
\label{eq:C5A} 
C^A_5(q^2) = C^A_5(0) \left(1-\frac{q^2}{M^2_{A\Delta}} \right)^{-2} \,.
\end{equation}
For it we have adopted a dipole ansatz with $C^A_5(0)=1.00\pm0.11$ and $M_A=0.93$~GeV, following the fit to $\nu_\mu d \to \mu^- \Delta^{++} n$ BNL and ANL data~\cite{Hernandez:2010bx}.

We have extended the model to higher energy transfers by taking into account intermediate states from  the second resonance region, which includes three isospin 1/2 baryon resonances $P_{11}(1440)$, $D_{13}(1520)$ and $S_{11}(1535)$. The structure of the contribution of $P_{11}(1440)$ and $S_{11}(1535)$ to the amputated amplitudes is similar to the one of the nucleon, while the $N(1520)$ amplitudes resemble the $\Delta(1232)$ ones. Detailed expressions can be found in Ref.~\cite{Wang:2013wva}. With these $N^*$ terms we have adopted a strategy similar to the one for the $\Delta(1232)$ terms. The vector form factors are expressed in terms of the empirical helicity amplitudes extracted in the MAID analysis. As there is no experimental information to constrain the axial form factors, following Ref.~\cite{Leitner:2008ue}, we have kept only the leading axial terms and used PCAC to derive off-diagonal Goldberger-Treiman relations between the corresponding axial couplings and the $N^* \to N \pi$ partial decay widths. For the $q^2$ dependence we have assumed a dipole ansatz like in Eq.~(\ref{eq:C5A}) with a natural value of  $M^*_A=1.0$ GeV. 

Finally, the $\pi Ex$ mechanism originates from the $Z\gamma\pi^0$ vertex fixed by the axial anomaly of QCD 
\be
\Gamma^{\mu \rho}_{\pi Ex}  =  eC_N\frac{g_A M}{4\pi^2 f_\pi^2}(1-4
\sin^2\theta_W) \frac{\epsilon^{\mu\rho\sigma\alpha}q_\sigma (k_{\gamma})_\alpha}{(q-k_\gamma)^2-m_\pi^2} \gamma_5, \qquad \left(C_p=+1\,,\, C_n=-1 \right) \,.
\label{eq:piex}
\ee
It is nominally of higher order~\cite{Serot:2012rd} and, indeed, gives a very small contribution to the cross section as will be shown below. There are other $t$-channel $\rho$ and $\omega$ exchange mechanisms~\cite{Hill:2009ek} that arise from the anomaly-mediated $Z^0\gamma \rho$ and $Z^0\gamma \omega$ interactions~\cite{Harvey:2007rd}. Among them, the $\omega$ contribution is favored by the size of the couplings. Reference~\cite{Serot:2012rd} assumes that the $\rho$ and $\omega$ exchange mechanisms, taken from Ref.~\cite{Hill:2009ek}, saturate the low-energy constants in the contact terms. The contribution of these contact terms to the NC$\gamma$ cross section on the nucleon is very small at $E_\nu \leq 550$~MeV~\cite{Hill:2009ek,Serot:2012rd}, as expected from power counting arguments. The extension to higher energies requires the introduction of poorly understood form factors~\cite{Hill:2009ek,Zhang:2012xn}. The cross section from these mechanisms increases fast with energy, which might be a concern for experiments at high energies, or with a high-energy tail in the neutrino flux (like T2K). Nevertheless one should recall that this trend will be limited by unitarity bounds: these amplitudes will be modified by loop contributions and partially canceled by contact terms of even higher orders. 

\subsection{NC$\gamma$ on nuclei}

This reaction on nuclear targets can be incoherent
\begin{equation}
  \nu,\bar\nu (k) +\, A_Z   \to \nu,\bar\nu (k^\prime) + \,
   \gamma(k_\gamma) +\, X,\qquad   
\label{eq:reacincl}
\end{equation}
or coherent 
\be
\nu,\bar\nu (k) +\, A_Z|_{gs}(p_A)  \to \nu,\bar\nu (k^\prime) +
  A_Z|_{gs}(p^\prime_A) +\, \gamma(k_\gamma)
\label{eq:reacoh}
\ee
depending on whether the final nucleus is in an excited or in its ground state.

\begin{figure}[h!]
\includegraphics[width=.43\textwidth]{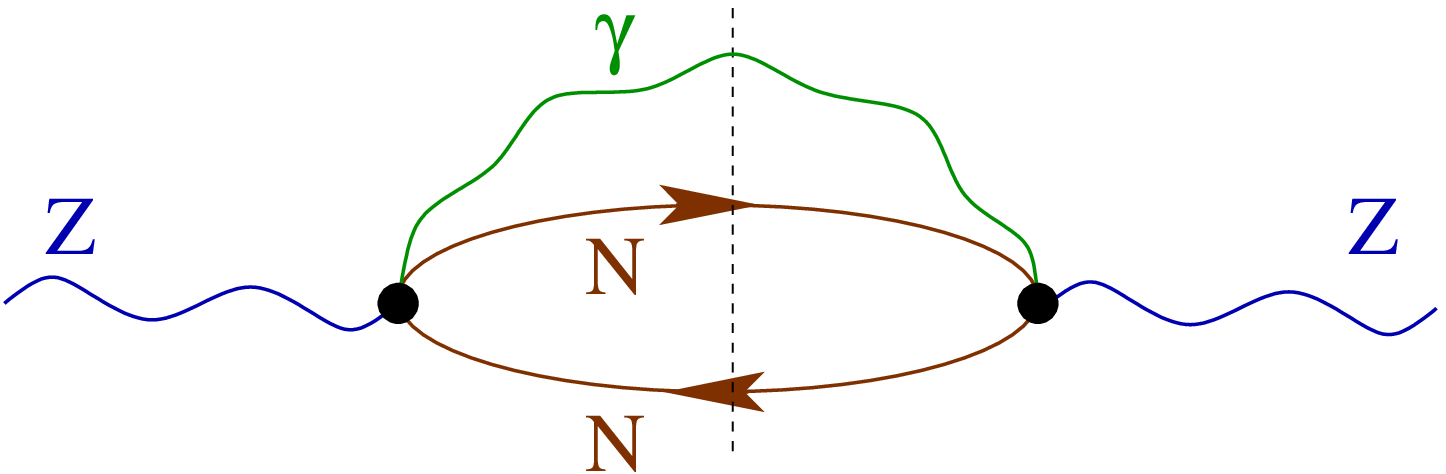}
\caption{Diagrammatic representation of the one-particle-one-hole-photon (1p1h$\gamma$) 
contributions to the $Z^0$ self-energy in nuclear matter. The black dots represent 
$Z^0N \to \gamma N$ elementary amplitudes. To obtain the imaginary part, the intermediate states intersected by the dashed line have to be placed on the mass shell.}\label{fig:1p1h}
\end{figure}

In a many-body scheme adapted to (semi)inclusive reactions on finite nuclei by means of  the local density approximation, the NC$\gamma$ incoherent cross section is 
\begin{equation}
\left. \sigma_{(\nu,\bar\nu)}\right|_{\mathrm{incoh}} =
    \frac{1}{ |\vec{k}~|}\frac{G^2}{16\pi^2}
     \int  \frac{d^3k'}{|\vec{k}^\prime|}
    L_{\mu\sigma}^{(\nu,\bar\nu)}\left. W^{\mu\sigma}_{{\rm NC}\gamma}\right|_{\mathrm{incoh}} \label{eq:sec-incl}\,.
\end{equation}
The hadronic tensor is obtained from the imaginary part of the contributions to the $Z^0$ selfenergy with a single photon in the intermediate state. In a density expansion, the lowest order contribution  is  depicted in Fig.~\ref{fig:1p1h}. 

After the simplification of evaluating the $\Gamma^{\mu\rho}_{N}$ amplitudes at an average nucleon hole
four momentum $\langle p^\mu \rangle$, $W^{\mu\sigma}_{{\rm NC}\gamma}$ becomes (see Ref.~\cite{Wang:2013wva} and references therein for details) 
\begin{equation}
W^{\mu\nu}_{1{\rm p1h}\gamma}(q) = \Theta(q^0) \frac{1}{2M^2} \int
\frac{d^3r }{2\pi} 
\sum_{N=p,n} \frac{d^3k_\gamma}{(2\pi)^3}
\frac{\Theta(q^0-E_\gamma)}{2E_\gamma} {\rm
  Im}\overline {U}_R(q-k_\gamma,k_F^N,k_F^{N}) A^{\nu\mu}_{N} \label{eq:1p1hga-def}
\end{equation}
where
\be
A^{\mu\nu}_{N} = \frac12 {\rm
Tr}\left[\left(\langle\slashchar{p}\rangle+M\right)\gamma^0 \left( \left\langle \Gamma_{N}  \right \rangle^{
  \mu \rho} \right)^\dagger \gamma^0\left(\langle\slashchar{p}\rangle+\slashchar{q}
-\slashchar{k}_\gamma+M\right) \left\langle \Gamma_{N}\right \rangle^\nu_{.\,\rho} \right]
\ee
while $\overline {U}_R(q-k_\gamma,k_F^N,k_F^{N})$ stands for the  Lindhard function (definition and explicit expressions can be found in Ref.~\cite{Nieves:2004wx}). The Fermi momentum depends on the local density of nucleons in the nucleus via $k^N_F(r)= [3\pi^2\rho_N(r)]^{1/3}$. The density distributions are based on empirical determinations in the case of protons and on realistic theoretical models in the case of neutrons.

It is known that the $\Delta(1232)$ properties are strongly modified in the nuclear medium. This important nuclear correction is taken into account by the following substitutions in the $\Delta(1232)$ propagator
\bea
M_\Delta &\raw& M_\Delta + \mathrm{Re} \Sigma_\Delta(\rho)\,, \\
\Gamma_\Delta &\raw& \tilde{\Gamma}_\Delta - 2\,  \mathrm{Im} \Sigma_\Delta(\rho)\,.
\eea
The real part of the in-medium $\Delta$ selfenergy, $\Sigma_\Delta$, receives an attractive (negative) contribution from the nuclear mean field, which is partially cancelled by an effective repulsive piece from iterated $\Delta$-hole excitations.  As the net effect is smaller than the precision 
achievable in current neutrino experiment we simply take  $\mathrm{Re} \Sigma_\Delta(\rho) \approx 0$. The resonance decay width is reduced to  $\tilde{\Gamma}_\Delta$ because the final nucleon in $\Delta \raw \pi N$ can be Pauli blocked but, on the other hand, it increases because of the presence of many body processes such as $\Delta \, N \raw N \, N$, $\Delta \, N \raw N \,N \, \pi$ and  $\Delta \, N \, N \raw N \, N \, N$. These new decay channels, accounted in $\mathrm{Im} \Sigma_\Delta$, have been parametrized as a function of the local density in Ref.~\cite{Oset:1987re}.

For the coherent reaction of Eq.~(\ref{eq:reacoh}) one has~\cite{Wang:2013wva,Amaro:2008hd} that 
\be
\left.\frac{d^{\,3}\sigma_{(\nu,\bar\nu)}}{dE_\gamma
  d\Omega(\hat{k}_\gamma)}\right|_{\rm coh}  =
    \frac{E_\gamma}{ |\vec{k}~|}\frac{G^2}{16\pi^2}
     \int  \frac{d^3k'}{|\vec{k}^\prime|}
    L_{\mu\sigma}^{(\nu,\bar\nu)} \left. W^{\mu\sigma}_{{\rm
        NC}\gamma}\right|_{\rm coh} 
\ee
with 
\be
\left. W^{\mu\sigma}_{{\rm NC}\gamma}\right|_{\rm coh} = -
  \frac{\delta(E_\gamma-q^0)}{64\pi^3M^2}  {\cal A}^{\mu\rho}(q,k_\gamma)
    \left({\cal A}^\sigma_{.\,\rho}\right)^*(q,k_\gamma) \label{eq:zmunu} \,,
\ee
\be
{\cal A}^{\mu\rho}(q,k_\gamma) = 
\int d^3r\ e^{i\left(\vec{q}-\vec{k}_\gamma\right)\cdot\vec{r}} 
\left\{\rho_{p}(r\,) {\hat \Gamma}^{\mu\rho}_{p}(r;q,k_\gamma) 
+ \rho_{n}(r\,)
    {\hat \Gamma}^{\mu\rho}_{n}(r;q,k_\gamma) \right\}\label{eq:Jmunu2} \,.
\ee
and
\be
{\hat \Gamma}^{\mu\rho}_{N}(r;q,k_\gamma) = 
\sum_i \frac12 {\rm Tr}\left[(\slashchar{p}+M)\gamma^0\,\Gamma_{i;
  N}^{\mu\rho} \right] \left. \frac{M}{p^0}  
  \right|_{p^\mu=\left( \sqrt{M^2+\frac{(\vec{k}_\gamma-\vec{q}\,)^2}{4}},
      \frac12(\vec{k}_\gamma-\vec{q}\,)\right)}  \,, 
\label{eq:cn-jcoh} 
\ee
where $\Gamma_{i; N\gamma}^{\mu\rho}$ stand for the amputated photon production amplitudes  for the different mechanisms $i=NP,\, CNP,\,\pi Ex,\, RP,\, CRP \,\, \left [R=\Delta,N(1440), N(1535), N(1520)\right ]$. In this case, the nucleon wave functions remain unchanged so that one has to sum over all nucleons at the amplitude (not amplitude squared) level. This leads to the trace in Eq.~(\ref{eq:cn-jcoh}) and the nuclear density distributions in  Eq.~(\ref{eq:Jmunu2}).  Therefore, the coherent production process is sensitive to the Fourier transform of the nuclear density.  In the elementary $Z^0 N \to N \gamma$ process, energy conservation is accomplished by imposing
$q^0=E_\gamma$, which is justified by the large nucleus mass. The transferred momentum is assumed to be equally shared between the initial and final nucleons. A lengthy discussion about this prescription and the local treatment of the $\Delta$ propagation can be found in Ref.~\cite{Wang:2013wva}. The modification of the $\Delta$ in the medium outlined above has also been considered here.

\section{NC$\gamma$ cross sections}

\subsection{On nucleons}

The integrated NC$\gamma$ cross sections on protons and neutrons as a function of the (anti)neutrino laboratory energy are displayed in Fig.~\ref{fig:NCgamma_nucleon}. As in other processes, the different helicities of $\nu$ and $\bar{\nu}$ are responsible for different interferences, resulting in smaller $\bar{\nu}$ cross sections with a more linear energy dependence. The $\Delta$ mechanism is
dominant. Its contribution is the same on protons and neutrons because of the isovector nature of the electroweak $N-\Delta$ transition. The contribution of $NP + CNP$ terms is also important, being only about 2.5 smaller than the $\Delta$ one at $E_{\nu(\bar \nu)} \sim 1.5$ GeV. Above $\sim 1.5$~GeV,
the $N(1520)$ contribution is also sizable and, for $\bar\nu$  on protons, comparable to the one of $NP + CNP$. However, the $N(1440)$ and $N(1535)$ and $\pi Ex$ mechanisms  can be safely neglected. The fact that the  $N(1520)$ resonance is the only one, besides the $\Delta(1232)$, playing a significant role at $E_\nu < 2$~GeV  has also been observed in pion production~\cite{Hernandez:2013jka} and for the inclusive cross section~\cite{Leitner:2008ue}. More details, including differential photon energy and angular distributions as well as comparisons with earlier studies can be found in Ref.~\cite{Wang:2013wva}.  
\bfig[h!]
\includegraphics[width=0.45\textwidth]{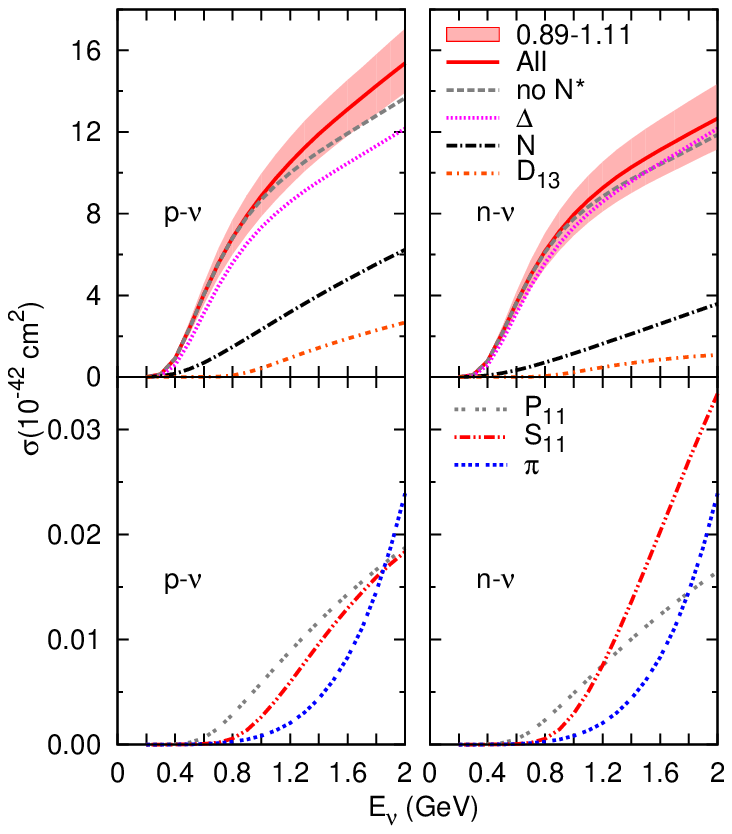}
\includegraphics[width=0.45\textwidth]{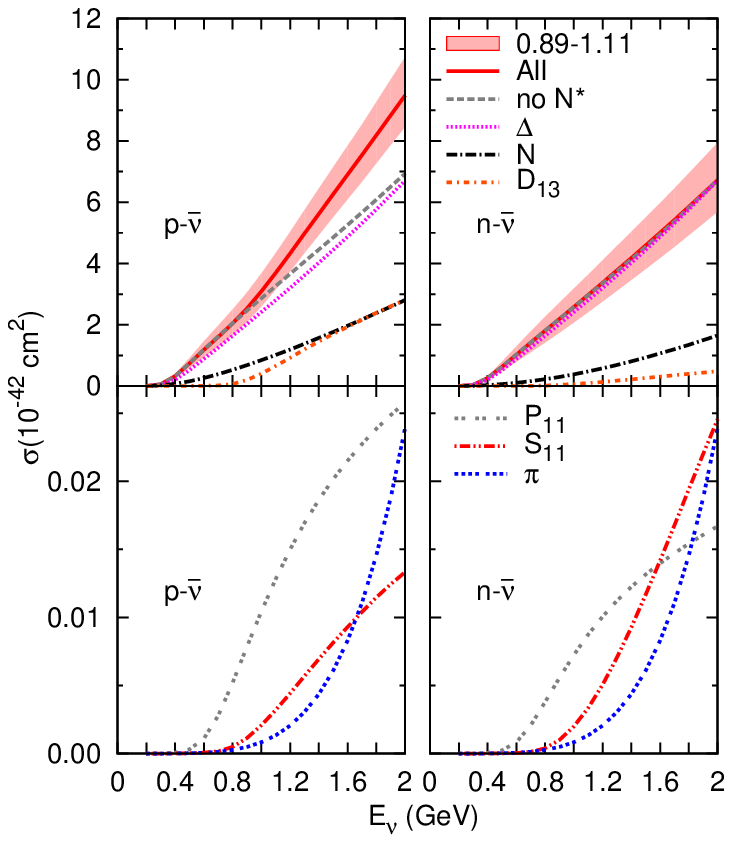}
\caption{\label{fig:NCgamma_nucleon} $\nu N \to \nu N \gamma$ (left panel) and $\bar\nu N
  \to \bar\nu N \gamma$ (right panel) cross sections on protons and neutrons~\cite{Wang:2013wva}. A cut of $E_\gamma \geq 140$~MeV in the phase space integrals has been applied. The error bands in the full-model results (solid lines) represent the uncertainty in the  axial $N\Delta$ coupling $C^A_5(0)=1.00 \pm 0.11$ 
according to the determination of Ref.~\cite{Hernandez:2010bx}. The curves labeled $N$, 
$\Delta$, $D_{13}$, $P_{11}$ and $S_{11}$ stand for the partial contributions of the $BP$ and $CBP$ mechanisms of Fig~\ref{fig:diags}; label $\pi$ corresponds to $\pi Ex$.
The $N^*$ contributions have been removed from the results labeled ``no $N^*$''. }
\efig

\subsection{On nuclei}

In the left panel of Fig.~\ref{fig:NCgamma_nucleus}, we show our predictions for the (anti)neutrino incoherent photon emission cross sections on $^{12}$C as a function of the (anti)neutrino energy. It is clear that neglecting nuclear medium corrections is a poor approximation. By taking into account Fermi motion and Pauli blocking, the cross section already goes down by more
than 10\%. With the full model that also includes 
the $\Delta$ resonance in-medium modification, the reduction is  of the
order of 30\%. 
\bfig[h!]
\includegraphics[width=0.45\textwidth]{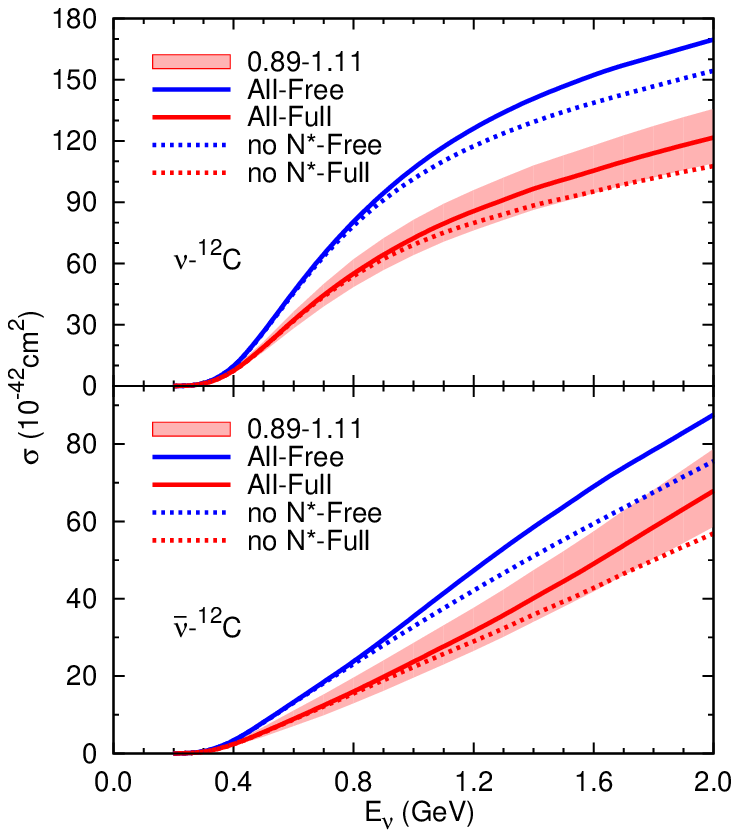}
\includegraphics[width=0.45\textwidth]{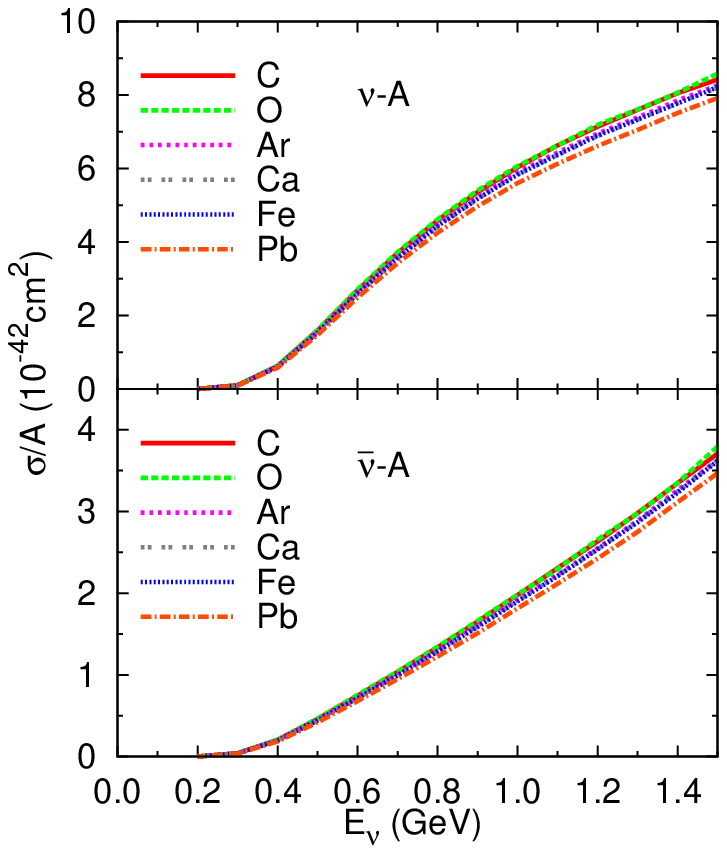}
\caption{\label{fig:NCgamma_nucleus} Left panel: Incoherent NC$\gamma$ cross sections on $^{12}$C according to the model of Ref.~\cite{Wang:2013wva}. 
All curves have been obtained with an $E_\gamma \geq 140$~MeV cut in the phase space. Solid lines are from the complete model at the nucleon level,  while the dotted lines were obtained without the $N^*$ contributions. Curves denoted as ``Free'' (upper blue curves) do
  not include any nuclear correction: ($\sigma_A=Z\sigma_p+N\sigma_n)$. 
 Curves labeled as ``Full'' (lower red curves) take into account nuclear effects. 
The error bands show the uncertainty from the the axial $N\Delta$ coupling 
($C^A_5(0) = 1.00 \pm 0.11$). Right panel: Integrated cross sections for  different nuclei ($^{12}$C,$^{16}$O,$^{40}$Ar, $^{40}$Ca,$^{56}$Fe and $^{208}$Pb) divided by the number of nucleons.}
\efig
The right panel of Fig.~\ref{fig:NCgamma_nucleus} shows total NC$\gamma$ incoherent cross
sections for different nuclei. The curves indicate an approximated $A$-scaling. Nevertheless, the cross section is smaller for heavier nuclei, particularly $^{208}$Pb. We should stress that the observed deviation from scaling  cannot be explained only by neutron cross sections being smaller than proton ones. Photon distributions do not show any appreciable $A$ dependence in the shapes as can be seen in Fig.~10 of Ref.~\cite{Wang:2013wva}. Comparisons with other calculations are also displayed and discussed in that article.

\begin{figure}[h!]
\includegraphics[width=0.45\textwidth]{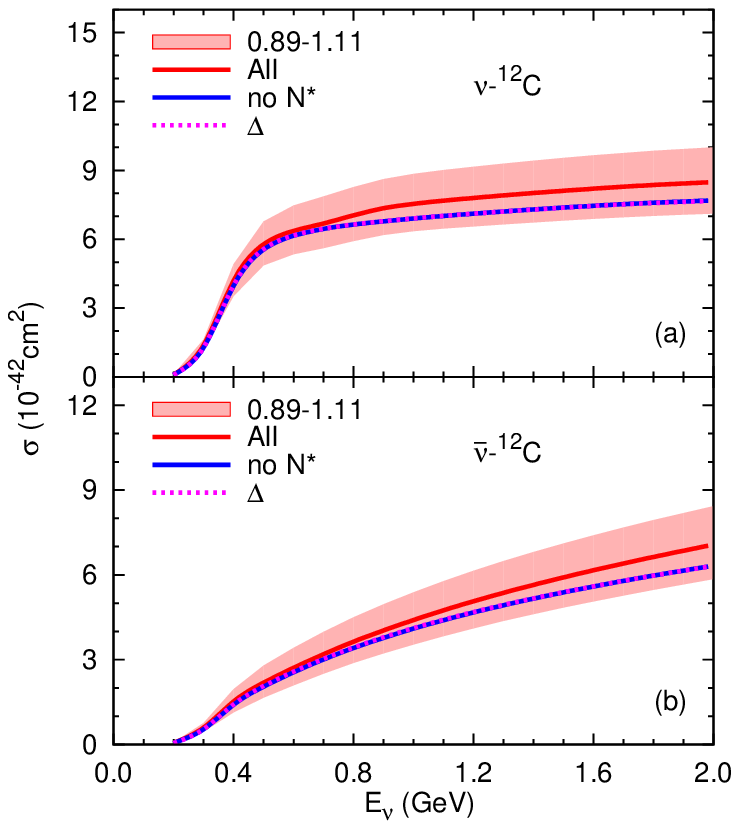}
\includegraphics[width=0.45\textwidth]{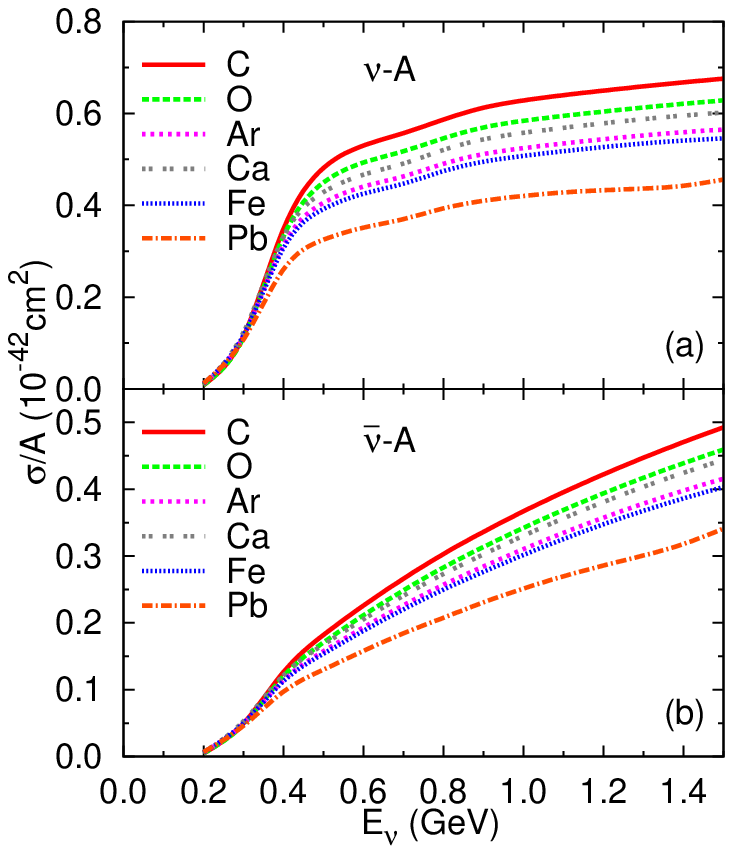}
\caption{Left panel:  Left panel: Coherent NC$\gamma$ cross sections on $^{12}$C according to the model of Ref.~\cite{Wang:2013wva}. A photon energy cut of  $E_\gamma \geq 140$~MeV has been applied. Red solid lines stand for results from the complete model, with error  bands determined by the uncertainty of $\pm 0.11$ in $C^A_5(0)$~\cite{Hernandez:2010bx}. The solid blue lines below, labeled as ``no $N^*$'', display the 
cross sections without the $N^*$ amplitudes, while the magenta dotted ones
denote the contributions from the $\Delta$ mechanisms alone. Right panel: Integrated cross sections for  different nuclei ($^{12}$C,$^{16}$O,$^{40}$Ar, $^{40}$Ca,$^{56}$Fe and $^{208}$Pb) divided by the number of nucleons. }
\label{fig:coh2}
\end{figure}

Integrated cross sections for coherent NC$\gamma$ are shown in Fig.~\ref{fig:coh2}. These are  about a factor 10-15 smaller than the incoherent ones. Thus, their relative relevance is similar if not greater than in pion production. As is apparent from the left panel, the dominance of the $ \Delta P+ C\Delta P$ terms is more pronounced in this case, with small corrections from $N(1520)$ excitation. Nucleon-pole contributions are negligible because the coherent kinematics favors a strong cancellation between the direct and crossed terms. The $\pi Ex$ terms vanish exactly for isospin symmetric nuclei because amplitudes for protons and neutrons cancel with each other. 

Unlike $\pi$ and $\rho$ $t$-exchange terms, the $\omega$ contribution does not vanish for symmetric nuclei because amplitudes on protons and neutrons add up. In Ref.~\cite{Hill:2009ek} it was found that the coherent NC$\gamma$ from $\omega$ exchange plays a sub-dominant role at $E_\nu \sim 1$~GeV, compared to naive estimates, being suppressed by form factors and recoil.  On the other hand, because of their strong energy dependence of the contact terms, the coherent cross section in Ref.~\cite{Zhang:2012xn} is dominated by contact terms for $E_{\nu,\bar\nu} > 650$~MeV.  However, as discussed after Eq.~(\ref{eq:piex}), these results are not only highly sensitive to unknown form factors but should also be constrained by unitarity.  

Finally, the right panel reveals that the nuclear dependence is stronger than for the incoherent channel. The coherent cross sections neither scale with $A$, like the incoherent one approximately does, nor with $A^2$ as one would expect from the dominant isoscalar $\Delta P$ mechanism. This is related to the structure of the axial and vector currents at $\vec{q} \approx \vec{k}_\gamma$, favored by the nuclear form factor: a more elaborated explanation can be found in Ref.~\cite{Wang:2013wva}.

\section{Single photon events at MiniBooNE, T2K and NOMAD}

\subsection{NC$\gamma$ and the low energy excess at MiniBooNE}

The MiniBooNE experiment, designed to explore the short-baseline $\bar\nu_\mu \to \bar\nu_e$ oscillations, has found an excess of electron-like events over the predicted background in both $\nu$ and $\bar \nu$ modes~\cite{AguilarArevalo:2008rc,Aguilar-Arevalo:2013pmq}. The excess is concentrated at $200 < E_\nu^{\mathrm{QE}} < 475$~MeV, where $E_\nu^{\mathrm{QE}}$ is the neutrino energy reconstructed assuming a charged-current quasielastic (CCQE) nature of the events. Recent studies indicate that explanation of this anomaly cannot reside in oscillations, even involving  sterile neutrinos~~\cite{Conrad:2012qt,Giunti:2013aea}. On the other hand, it could have its origin in poorly understood backgrounds.

At low $E_\nu^{\mathrm{QE}}$ the background is dominated by photon emission because Cherenkov detectors like MiniBooNE cannot distinguish electrons from single photons. The largest source of single photons is NC$\pi^0$ production where one of the photons from the $\pi^0 \raw \gamma \gamma$ decay is not identified. This background has been constrained by the MiniBooNE's NC$\pi^0$ measurement~\cite{AguilarArevalo:2009ww}. The second most important process is NC$\gamma$. The MiniBooNE analysis estimated this background using the NC$\pi^0$ measurement, assuming that NC$\gamma$ events come from the radiative decay of weakly produced resonances, mainly $\Delta \to N \gamma$~\cite{AguilarArevalo:2008rc,Aguilar-Arevalo:2013pmq}. This procedure neither takes into account the existence of non-resonant terms, nor the coherent channel. If the NC$\gamma$ emission estimate were not sufficiently accurate, this would be relevant to track the origin of the observed excess. 

We have applied the model described above to calculate the number and distributions of single photon events at MiniBooNE~\cite{Wang:2014nat}, using the available information about the detector mass (806 tons) and composition (CH$_2$), the total number of protons on target (POT), $6.46 \times 10^{20}$ in $\nu$ mode and $11.27 \times 10^{20}$ in $\bar{\nu}$ mode,~\cite{Aguilar-Arevalo:2013pmq}, the flux prediction~\cite{AguilarArevalo:2008yp} and photon detection efficiency~\cite{mbweb}. The yields from the incoherent channel are the largest ones. Those from the coherent channel and the reaction on protons, which are comparable, are smaller but significant. The coherent contribution is particularly important for antineutrinos and in the forward direction (see the plots in Ref.~\cite{Wang:2014nat}).    

\begin{figure}[h!]
\includegraphics[width=0.45\textwidth]{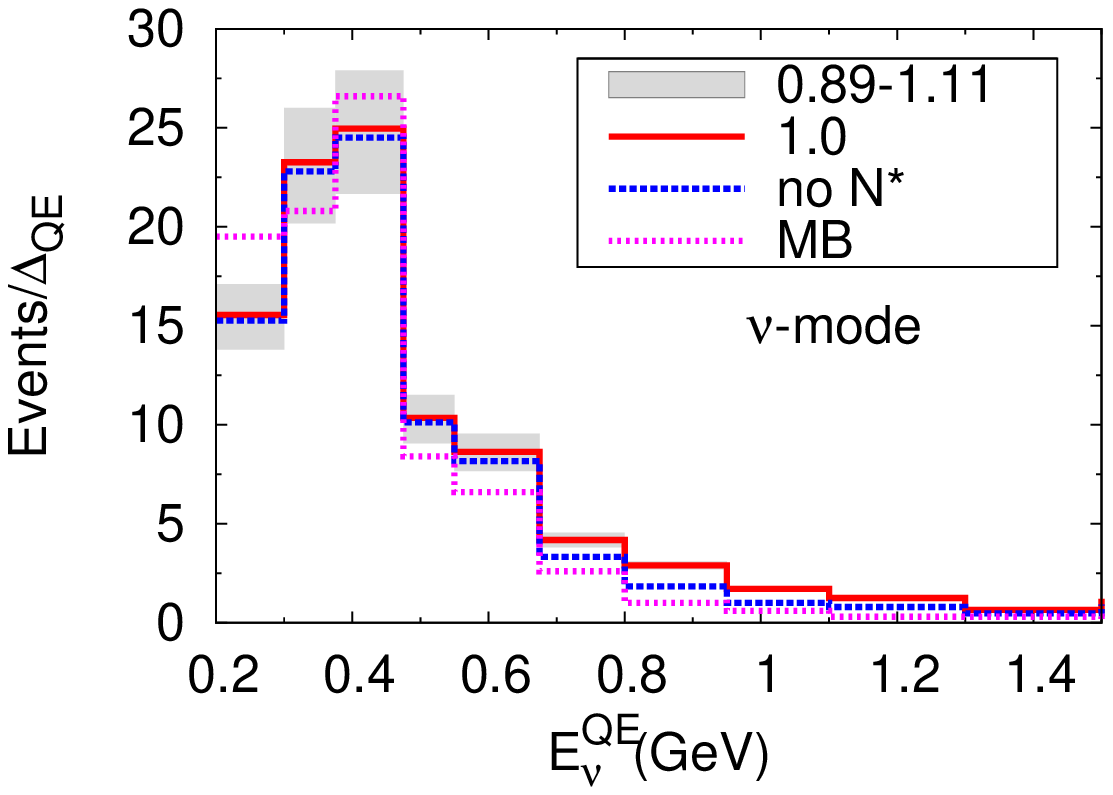}
\includegraphics[width=0.45\textwidth]{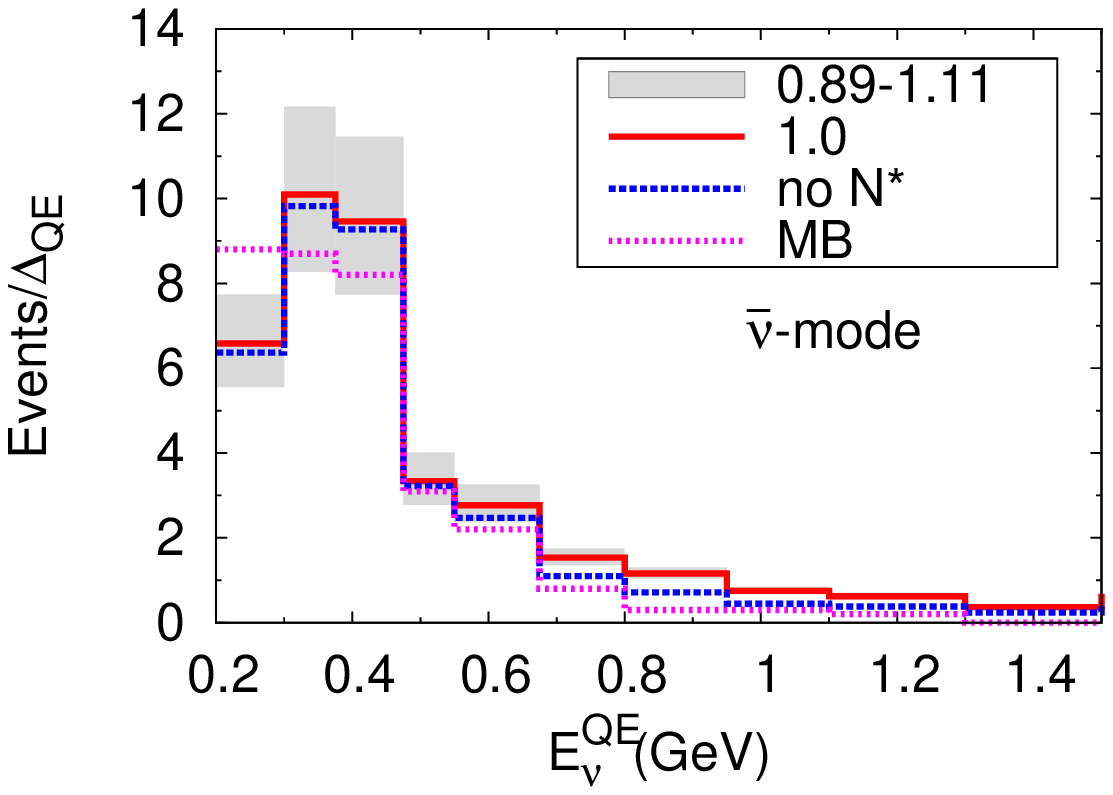}  
\caption{
\label{fig:re}
 $E^\mathrm{QE}_\nu$ distributions of total  NC$\gamma$  events 
for the $\nu$ (left) and $\bar \nu$ (right) modes. The
 error bands correspond to the uncertainty in $C^A_5(0)=1.00\pm
 0.11$~\cite{Hernandez:2010bx} around the central value labeled as ``1.0''.
The curves labeled as ``no $N^*$'' show results without the 
$N^*$ contributions. The ``MB'' histograms display the MiniBooNE estimates~\cite{mbweb}.}
\end{figure}

Our results for the $E^\mathrm{QE}_{\nu}$ distributions are shown in Fig.~\ref{fig:re}. The error bands correspond to the uncertainty in $C^A_5(0)$. A more complete error analysis, leading to similar bands, can be found in Ref.~\cite{Wang:2014nat}.  The comparison with the MiniBooNE in situ estimate~\cite{Aguilar-Arevalo:2013pmq,mbweb} shows a good agreement: the shapes are similar and the peak positions coincide. The largest discrepancy is observed in the lowest energy bin. In the two bins with the largest number of events, the two calculations are consistent within our errorbars. For higher $E^\mathrm{QE}_\nu$ values, our results are systematically above the MiniBooNE estimate although the differences are very small. The error in the detection efficiency ($\sim 15\%$)~\cite{mbweb}, not considered in this comparison, will partially account for the discrepancies.
The overall agreement is also good in comparison to the estimate of Zhang and Serot~\cite{Zhang:2012xn} in spite of the differences in the approaches, in contrast to the findings of Hill~\cite{Hill:2010zy}, obtained with a rather high and energy independent detection efficiency, and neglecting nuclear effects. Therefore, we have found that neutral current photon emission from single-nucleon currents is insufficient to explain the event excess observed by MiniBooNE in both neutrino and antineutrino modes. 

\subsection{NC$\gamma$ events at the Super-Kamiokande detector}

The T2K experiment has measured $\nu_e$ appearance in a $\nu_\mu$ beam, obtaining the first indication of a nonzero value of $\theta_{13}$~\cite{Abe:2011sj}. The increasing precision allows for more detailed studies of neutrino properties and might lead to the discovery of $CP$ violation in the lepton sector. Progress in this direction requires a better control over irreducible backgrounds. As MiniBooNE, the Super-Kamiokande (SK) water detector is a Cherenkov one and cannot distinguish photons from electrons. The largest part of this background comes from NC$\pi^0$, which has been significantly reduced using specific reconstruction algorithms~\cite{Abe:2013hdq}, making the irreducible NC$\gamma$ contribution relatively more important.

With our model we have predicted the number and distributions of NC$\gamma$ events at the SK detector. We have used the flux of the off-axis neutrino beam from Tokai at SK~\cite{Abe:2012av} but neglect the tail above $E_\nu =3$~GeV. For the recent T2K $\nu_e$ appearance analysis, corresponding to $N_{\mathrm{POT}} =  6.57 \times 10^{20}$ in $\nu$ mode~\cite{Abe:2013hdq} we have obtained a total number of 
\begin{equation}
{\cal N} = 0.421 \pm 0.051  \,,
\ee 
before efficiency corrections, with the error corresponding to the uncertainty in $C^A_5(0)$. When compared to the equivalent calculation performed with the NEUT event generator~\cite{Hayato:2009zz}, our result turns out to be 2-3 times larger. The disagreement is mostly in the normalization because the shapes of the photon energy and angular distributions are similar~\cite{shimpei}. This finding is in line with the comparison of the NC$\gamma$ integrated cross sections on $^{12}$C from different models presented in Fig.~9 of Ref.~\cite{Katori:2014qta}, where the NEUT result is below the rest.

\subsection{The NC$\gamma$ limit at NOMAD}

The high resolution data taken by the NOMAD experiment allow a sensitive search for
neutrino induced single photon events. The experiment has obtained an upper limit of $4.0 \times 10^{-4}$ single photon events per $\nu_\mu$ charged-current ones with 90~\% CL, at $E_\nu \sim 25$~GeV~\cite{Kullenberg:2011rd}. Althogh the NC$\gamma$  models developed so far are not applicable at the high energy transfers that can occur in NOMAD, in the limited region of phase space where these models are valid, they should fulfil the NOMAD constraint as a necessary condition. In our case, restricting the invariant mass of the outgoing nucleon-photon pair to $W < 1.6$~GeV, where the model can be retained applicable, and neglecting nuclear effects (that would reduce the cross section) for simplicity, 
we obtain 
\be
\frac{\sigma(\mathrm{NC}\gamma, W <1.6\,\,\mathrm{GeV})}{\sigma(\nu_\mu A \rightarrow \mu^- X)} \approx 0.8  \times 10^{-4} 
\ee
at  $E_\nu = 25$~GeV using NOMAD inclusive charged current cross section data~\cite{Wu:2007ab} for the denominator. Our prediction is then safely below the NOMAD limit.

\section{Outlook}

With the microscopic model developed in Ref.~\cite{Wang:2013wva} we have calculated photon emission on nucleons and nuclei in a kinematic region of interest for current and future neutrino experiments. After the inclusion of $N^*$ excitation mechanisms from the three lightest states, the model can be considered reliable up to nucleon-photon invariant masses of around 1.6~GeV. The main uncertainty arises from the uncertainty in the dominant $N-\Delta(1232)$ axial coupling that can be constrained in weak pion production experiments.  In the case of nuclear targets, we have also studied the mass dependence of both incoherent and coherent reaction channels emphazising the large ($\sim 30\%$) nuclear corrections.

We have predicted single photon events at the MiniBooNE and SK detectors where they represent an important electron-like irreducible background component. In the case of MiniBooNE, our results are consistent with the in-situ estimate, obtained with a much poorer model. Based on this, we conclude that photon emission processes from single-nucleon currents cannot  explain the excess of the signal-like events observed at MiniBooNE. 
 \begin{figure}[h!]
\includegraphics[width=.43\textwidth]{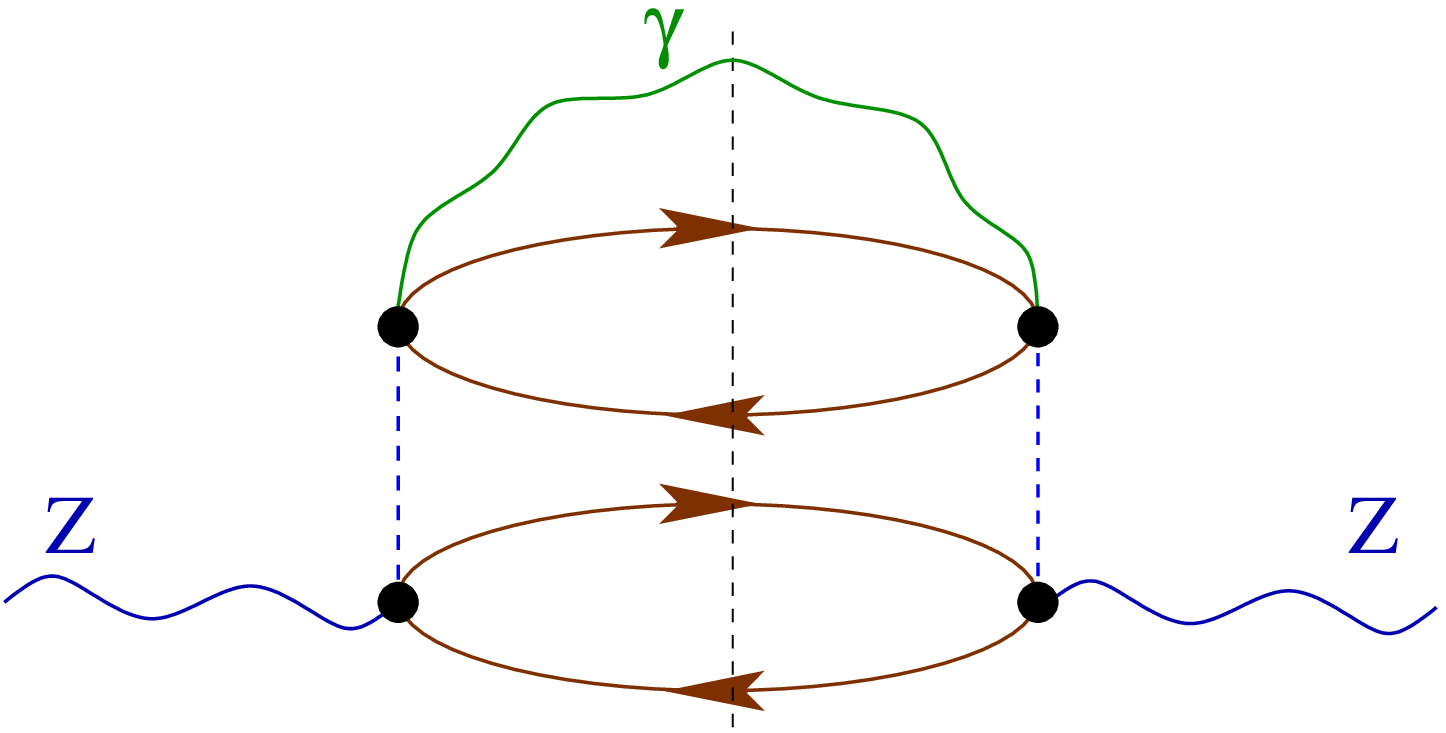}
\caption{Diagrammatic representation of two-particle-two-hole-photon (2p2h$\gamma$) 
contributions to the $Z^0$ self-energy in nuclear matter. The black dots represent 
$Z^0N \to \gamma N$ elementary amplitudes and the dash lines denote strong interactions.
To obtain the immaginary part, the intermedate states intersected by the dashed line have to be placed on the mass shell.}\label{fig:2p2h1gamma}
\end{figure}
As mentioned above, explanations of the MiniBooNE excess of events bases on oscillations do not seem  very plausible. Multinucleon mechanisms like those in Fig.~\ref{fig:2p2h1gamma} are being investigated. There are other possible explanations, like the possibility that a heavy neutrino is produced by weak~\cite{Gninenko:2009ks} or electromagnetic~\cite{Masip:2012ke} interactions in the detector, decaying radiatively afterwards. Such scenarios can be investigated in the forthcoming MicroBooNE experiment, capable of distinguishing photons from electrons. It should be recalled that any explanation of the MiniBooNE anomaly in terms of single photons, using the physics of the Standard Model or beyond it, should satisfy the NOMAD upper limit.

\begin{theacknowledgments}
Our work has benefited from discussions and useful communications by Y. Hayato, T. Katori, K. Mahn, P. Masjuan, S. Mishra, H. Tanaka, S. Tobayama and G. Zeller. Research supported by the Spanish Ministerio de Econom\'\i a y Competitividad and European FEDER funds under contract FIS2011-28853-C02-02, the Spanish Consolider-Ingenio 2010 Program CPAN (CSD2007-00042), Generalitat Valenciana under contract PROMETEO/2009/0090 and by the EU HadronPhysics3 project, grant agreement no. 283286. The work of LAR has been partially supported by the U.S. Department of Energy, Office of Science, Office of High Energy Physics, through the Fermilab Intensity Frontier Fellows Program. He gratefully acknowledges the hospitality during his stay at Fermilab.
\end{theacknowledgments}

\bibliographystyle{aipproc}   

\bibliography{neutrinos}

\end{document}